\documentclass[aps,nolongbibliography,superscriptaddress,reprint,10pt]{revtex4-2}

\usepackage{siunitx}
\DeclareSIUnit{\angstrom}{\textup{\r A}}

\usepackage{graphicx}
\usepackage{xcolor}
\usepackage[normalem]{ulem}
\usepackage{bm}

\newcommand{\textsub}[2]{{#1}_{\text{#2}}}
\newcommand{\EF}{E_{\text F}}
\newcommand{\MgGaic}{(MgGa)$_{\text{ic}}$}
\newcommand{\MgGaIC}[1]{(MgGa)$_{\text{ic}}^{#1}$}
\newcommand{\MgGaGai}{Mg$_{\text{Ga}}$Ga$_{\text i}$}

\newcommand{\EQ}[1]{Eq.\,(\ref{#1})}
\newcommand{\FIG}[1]{Fig.\,\ref{#1}}
\newcommand{\TBL}[1]{Table \ref{#1}}

\begin{document}

\title{First-Principles Study of Recombination-Enhanced Migration of an Interstitial Magnesium in Gallium Nitride}

\author{Yuansheng Zhao}
\affiliation{Institute of Materials and Systems for Sustainability, Nagoya University, Nagoya 464-8603, Japan}
\email{zhao.yuansheng.u3@f.mail.nagoya-u.ac.jp}

\author{Kenji Shiraishi}
\affiliation{Institute of Materials and Systems for Sustainability, Nagoya University, Nagoya 464-8603, Japan}
\affiliation{Graduate School of Engineering, Nagoya University, Nagoya 464-8601, Japan}

\author{Tetsuo Narita}
\affiliation{Toyota Central R\&D Labs., Inc., Nagakute, Aichi 480-1192, Japan}

\author{Atsushi Oshiyama}
\affiliation{Institute of Materials and Systems for Sustainability, Nagoya University, Nagoya 464-8603, Japan}

\begin{abstract}
The stable and metastable configurations of interstitial Mg in GaN and its migration energy barriers are studied from first-principles calculations.
In addition to the conventional octahedral (O, global energy minimum) and tetrahedral (T, metastable) interstitial sites, we have discovered two metastable interstitial complexes with formation energy lower than or close to that of T configuration but higher than O.
Except for Mg at O site which only has $+2$ charge state, all other configurations also permit charge states $+1$ or $0$.
The minimum migration energy barrier for Mg$^{++}$ between O sites is found to be \SI{1.95}{eV}.
We further find that, when Fermi energy is close to the conduction band, the migration between O sites via metastable configurations occurs through a recombination-enhanced mechanism in which the charge state changes from $+2$ at O site to $0$ at metastable sites by consecutive capture of two electrons during the migration.
This process greatly reduces the migration energy barrier to as low as \SI{1.47}{eV}. This value is consistent with experiments, and we also discuss the role of intrinsic defects in the migration of Mg.
\end{abstract}

\maketitle

Doping of donor and acceptor impurities which generate electrons and holes carrying electric currents is essential in the fabrication of electronic devices, thus being a central issue in semiconductor science and technology \cite{sze2021physics,narita2020characterization,pantelides1978electronic}. 
The doping can be fulfilled during epitaxial growth of semiconductor thin films. However, to achieve the desired profiles of donors and acceptors in complicated device structures, ion implantation followed by thermal annealing is indispensable. During such doping processes, atomic diffusion and migration are key phenomena which govern the profiles of donor and acceptor concentrations. 

Gallium nitride (GaN) is a premier semiconductor in optoelectronics and promising to advance future power electronics. In GaN, an Mg atom substitutes for the Ga site and works as an acceptor exclusively. In the fabrication of planar optical devices, the Mg has been usually doped during the epitaxial growth of GaN \cite{akasaki2015nobel,amano2015nobel,nakamura2015nobel}. In the case of power devices, however, the Mg doping via the ion implantation technique is necessary but has not been achieved before due to N desorption during annealing. 
Recently, Sakurai and his collaborators at Nagoya University have succeeded in forging p-type GaN through Mg implantation followed by annealing under high pressure of nitrogen \cite{sakurai2019highly,sakurai2020redistribution}, advancing the doping technology in GaN to a next stage comparable to Si technology. At this point, unveiling the migration mechanism of an interstitial Mg atom which is generated in the implantation process in GaN becomes demanding. 

Atomistic knowledge as to the Mg migration in GaN is surprisingly poor. The only \textit{ab initio}
work is done by Miceli and Pasquarello \cite{miceli2017migration}. They have performed microscopic calculations based on the density functional theory (DFT) \cite{hohenberg1964inhomogeneous,kohn1965self} with generalized gradient approximation (GGA) \cite{perdew1996generalized}, assuming that the diffusing species is doubly positive magnesium Mg$^{++}$.
They have identified two (meta)stable interstitial sites, the octahedral (O) and tetrahedral (T) sites for Mg$^{++}$, and found three distinct pathways in the interstitial channels. The calculated migration barrier is more than \SI2{eV}. 
Due to the difficulty in identifying diffusion mechanisms for Mg in GaN, which depend on the situations, the experimental values for diffusion coefficients or energy barriers vary largely \cite{pan1999doping,porowski2002annealing,itoh2022substitutional}.
Recently, the barrier of interstitial migration is estimated to be $1.3\sim\SI{2.3}{eV}$ from $\beta^-$ emission channeling \cite{wahl2017lattice,wahl2021lattice}.
Therefore, the calculated value above is somewhat large and there should be a hidden mechanism for the Mg migration. 

In this Letter, we perform the first-principles calculations based on DFT for the interstitial Mg in GaN and find that, in addition to Mg$^{++}$, other charge states Mg$^{+}$ and Mg$^{0}$ appear during the migration. This variation of the charge states reduces the migration barrier by around \SI{0.5}{eV} (recombination-enhanced migration), to a value consistent with the experiments. We also find metastable complexes of Mg and Ga in the interstitial region which enrich the migration pathways. Roles of intrinsic defects presumably generated during the ion implantation for the Mg migration are also discussed.  

First-principles DFT calculations have been performed with the VASP package \cite{kresse1996efficient,kresse1999ultrasoft} using the projector-augmented potentials \cite{blochl1994projector} with Ga $3d$ states in the core \footnote{We have tested that including the $3d$ electrons affects the formation energy typically by less than \SI{0.1}{eV}.}. GGA \cite{perdew1996generalized} for the geometry optimization and hybrid approximation (HSE) \cite{heyd2003hybrid} for the electronic-structure and the total-energy calculations for the optimized geometries, are used to the exchange-correlation energy.
We set the cutoff energy of \SI{400}{eV} for the plane-wave basis and use the calculated values for lattice constants $a=\SI{3.23}{\angstrom}$ and $c=\SI{5.25}{\angstrom}$ which agree with the experimental values with an error of $\sim 1\%$. 
The calculated band gap of crystalline GaN is \SI{3.4}{eV} in the present hybrid approximation with the mixing parameter $\alpha=0.34$ of Fock exchange.
An Mg interstitial is embedded in a $4\times 4\times 3$ supercell and is optimized using $2\times 2\times 2$ $\bm k$-point mesh until the force acting on each atom is smaller than \SI{0.01}{eV\cdot \angstrom^{-1}}. 
For determination of migration pathways between the stable and metastable configurations, we use the nudged elastic band (NEB) method with improved tangential estimate \cite{henkelman2000improved}.

The formation energy of the geometric configuration $\zeta$ with charge state $q$ is computed as
\begin{eqnarray}\label{eq:form-erg}
    \Delta E_\zeta(q) & = & E_\zeta(q) - \left(\textsub E{ref} + \sum_i \Delta N_i\mu_i\right) \\ \nonumber 
    & & {}+ q(\textsub E{VBT}+\EF)+\textsub E{corr}(q).
\end{eqnarray}
Here, $E_\zeta(q)$ and $\textsub E{ref}$ are the total energies of the supercell models for the Mg interstitial and the reference structure, respectively; $\mu_i$ and $\Delta N_i$ are the chemical potential of the element $i$ and its difference in number from the reference structure, respectively; $\textsub E{VBT}$ and $\EF$ are the energy of valence band top (VBT) and Fermi-level position relative to VBT, respectively; $\textsub E{corr}(q)$ is the correction term for charge state $q$ due to the finite size of the supercell model \cite{freysoldt2009fully}.

\begin{figure}
    \includegraphics[scale=.63]{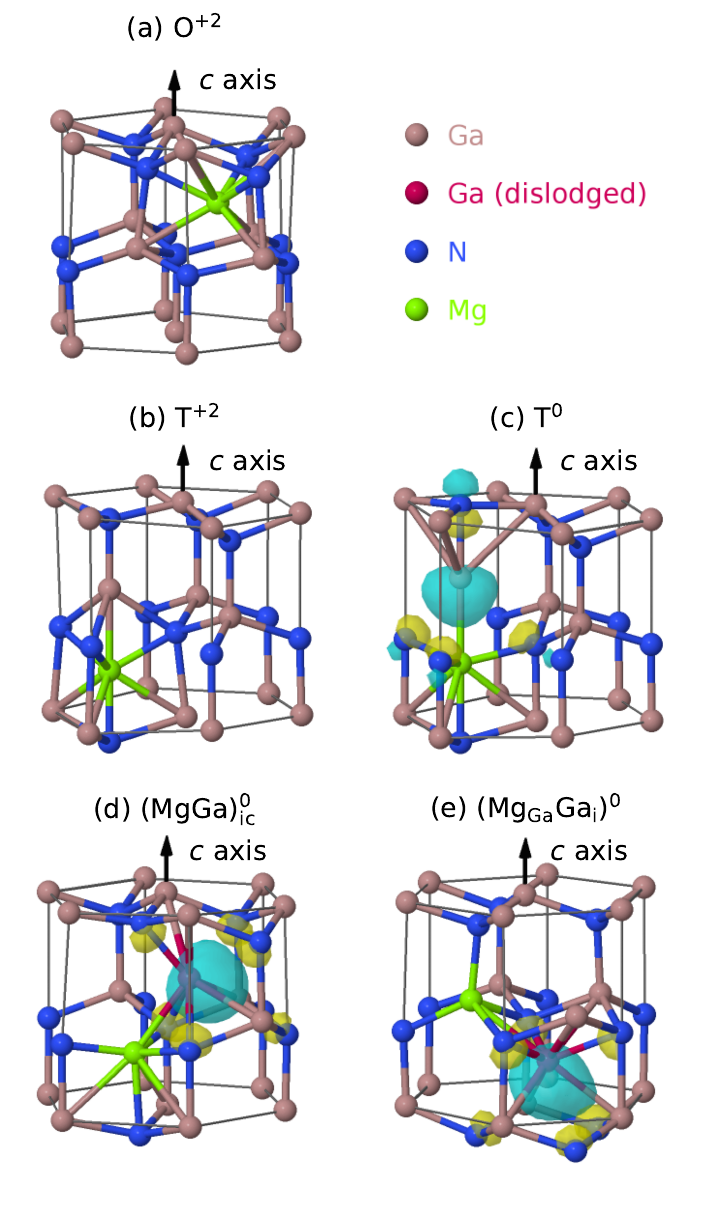}
    \caption{Stable and metastable configurations of interstitial Mg in GaN: 
    \textbf{(a)} O$^{+2}$, 
    \textbf{(b)} T$^{+2}$, 
    \textbf{(c)} T$^0$, 
    \textbf{(d)} \MgGaIC 0, and 
    \textbf{(e)} (\MgGaGai)$^0$ configurations. The superscripts denote the charge state. For the last three structures, the wavefunction of the gap state is also shown.
    }\label{fg:stable-sites}
\end{figure}

\begin{figure}
    \includegraphics[scale=.63]{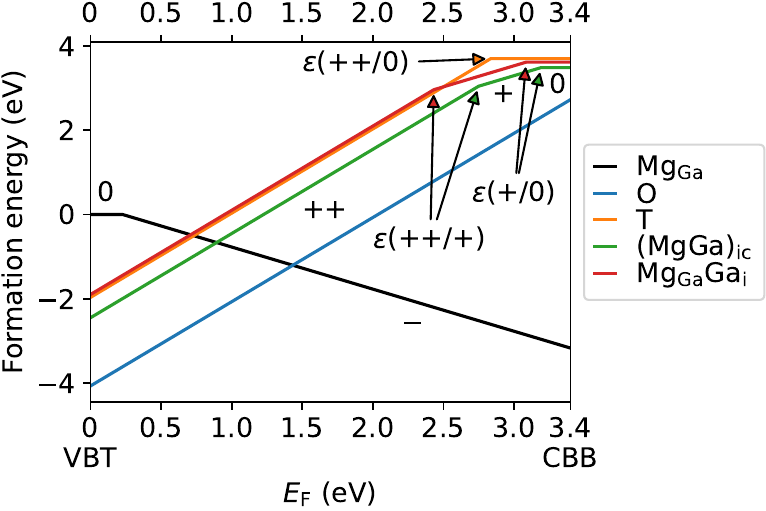}
    \caption{Formation energy of the interstitial Mg at various sites and with various charge states $q$ as a function of the Fermi-level position in the gap relative to that of the neutral substitutional Mg, Mg$_\text{Ga}$, under Ga-rich condition. The thermodynamic level $\varepsilon (q/q')$ is determined by the cross-point of the formation energies with $q$ and $q'$.}\label{fg:formation-erg}
\end{figure}

We set out with stable and metastable geometric configurations of an interstitial Mg in GaN by an extensive search with GGA calculations. 
The octahedral (O) and tetrahedral (T) sites are typical interstitial sites with high symmetry in wurtzite structure and previous DFT calculations \cite{miceli2017migration} examined their stability for doubly positive Mg. 
We have also examined the stability of Mg$^{++}$ at the O and T sites and found that both sites are stable although Mg is slightly dislodged from the high-symmetry points in both cases, as shown in \FIG{fg:stable-sites} (a), (b) and Table.\,SI in the Supplementary Material SM
\footnote{See Supplemental Material at [URL by publisher] for details of the calculations, and additional figures and tables.}. The O site is lower in the formation energy than the T site by 
\SI{2.10}{eV} which agrees well with the previous result \cite{miceli2017migration}. 
We have also examined charge states other than doubly positive. At the O site we have found that $+2$ is the only possible charge state. This is because Mg at the O site induces no gap state so that 2 additional valence electrons of Mg are missing for any position of the $\EF$ in the gap. In contrast, we have found that Mg at the T site becomes neutral when $\EF$ is near to the conduction band bottom (CBB), as is shown in \FIG{fg:formation-erg}. The formation energy is reduced by \SI{1.13}{eV} with Mg$_{\text T}^{++}$ becoming Mg$_{\text T}^{0}$ when $\EF$ is at CBB. The appearance of the neutral charge state at the T site is associated with substantial relaxation of surrounding atoms which induces the mid gap state [\FIG{fg:stable-sites} (c)].
Providing a single electron is not enough to attain such configuration. We have indeed found that the positively charged Mg at the T site is unstable. 

In addition to the Mg at the O and T sites, we have found two other metastable configurations which have never been addressed before: One is an interstitial complex, \MgGaic, a kind of ``split interstitial'', where Mg and dislodged Ga are located near a single lattice site with $\overrightarrow{\text{MgGa}}$ direction approximately parallel to $[4\bar 405]$, and the other is a pair of substitutional Mg and an interstitial Ga near the O site, \MgGaGai, as shown in \FIG{fg:stable-sites} (d) and (e), respectively. 
It is found that for both structures, the Mg and dislodged Ga both lie in $(1\bar 100)$ plane (Fig.\,S1 in the Supplementary Material SM \cite{Note2}).
For doubly positive charge state, the formation energies of \MgGaic{} and \MgGaGai{} are comparable with that at the T site (\FIG{fg:formation-erg}). 
The \MgGaic{} is actually found to be even lower than the T. 
More importantly, those configurations induce gap states which can capture electrons, thus rendering the $+1$ and neutral states realized. As shown in \FIG{fg:formation-erg}, the $+1$ and neutral \MgGaic{} and \MgGaGai{} appear when the $\EF$ is close to CBB. 
Yet we have found that, unlike the T configuration, the local structures are essentially the same for different charge states.

Several metastable configurations with distinct charge state we have found here open a possibility of carrier recombined migration of the interstitial Mg, in particular when $\EF$ is close to CBB.
The charge states other than $+2$ for Mg interstitial has never been discussed before. 
However, previous DFT calculations \cite{lyons2017computationally} show that intrinsic defects such as N vacancy or Ga interstitial (unavoidable during the ion implantation in device fabrication) induce donor-type gap levels. This may raise the Fermi-level position to near the conduction band and thus the metastable configurations with $q=+1$ or $0$ can become active.

\begin{figure}
    \includegraphics[scale=.63]{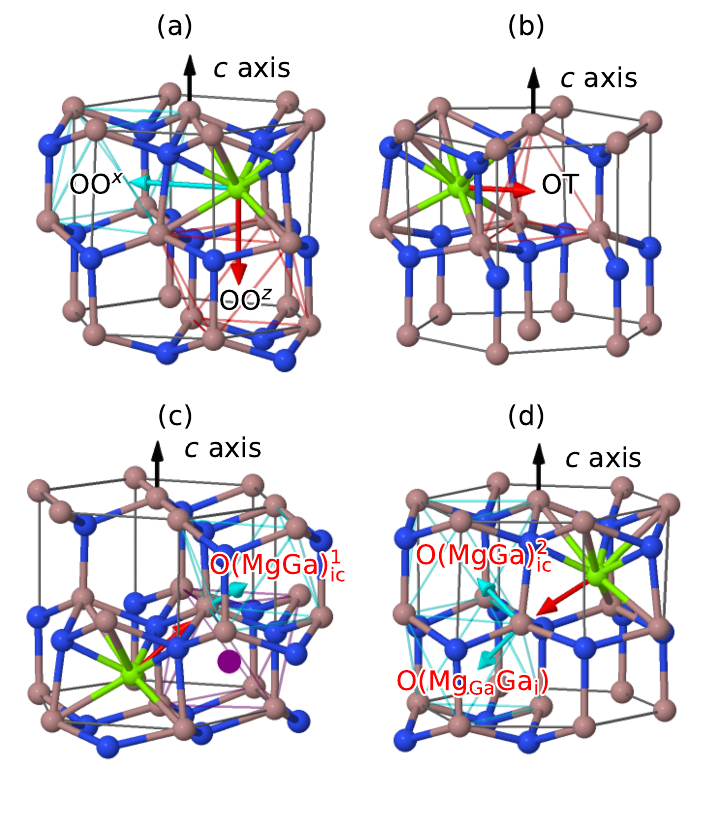}
    \caption{\textbf{(a)} OO$^z$ and OO$^x$ paths;
    \textbf{(b)} OT path;
    \textbf{(c)} O\MgGaIC 1 path, the red and light blue arrows show the movement of Mg and Ga, respectively; and
    \textbf{(d)} O\MgGaIC 2 path and O(\MgGaGai) path, the movement of Mg is the same in these two paths.
    }\label{fg:path}
\end{figure}

From the calculated formation energies, the interstitial Mg at the O site with +2 charge state is energetically most favorable irrespective of the $\EF$ position. Hence, the movement 
from one O site to a neighboring equivalent O site is the elementary process of the migration of the interstitial Mg. We have explored pathways for such elementary processes through NEB calculations. In the simplest pathways, 
as shown in \FIG{fg:path} (a), the Mg at an O site can migrate directly to nearby O site in $z$ direction ($c$ axis) or $xy$ plane without passing through any metastable sites. We refer to these two pathways as OO$^z$ and OO$^x$, respectively.
Clearly, the migration from an O site to one of the metastable sites and then to neighboring O site is also possible.
Due to structural symmetry, it suffices to examine the half of the pathways, i.e. the pathways between O and metastable sites. We have found 4 such pathways, as shown in \FIG{fg:path} (b)--(d). In the OT pathway, the Mg simply moves in $xy$ plane and migrates to a neighboring T site [\FIG{fg:path} (b)]. 
Regarding the pathways to \MgGaic{} or \MgGaGai, we have found 2 distinct processes, O\MgGaIC 1 and O\MgGaIC 2 (the superscripts denote the labels of pathways) where the moving directions of Mg are different as shown in \FIG{fg:path} (c) and (d), and a single process O(\MgGaGai) [\FIG{fg:path} (d)]. 
In these processes, the Mg at the O site moves towards one of the neighboring Ga atoms (red arrows in the figure) and pushes it to a nearby interstitial O site (light blue arrows) to form \MgGaic{} or \MgGaGai{} configuration. 
We emphasize that from the metastable sites, the Mg can go to \emph{another} nearby O site with the same energy barrier and continue to migrate.
For example, the O\MgGaIC 1 process followed by its reverse process \MgGaic O$^1$ can send the Mg to a neighboring O site, 
as is shown by the purple dot in \FIG{fg:path} (c), thus the combined processes are equivalent to an OO$^x$ process. 
In this way, the Mg movement between the O site and the metastable configuration effectively contributes to the Mg migration between the most stable O sites.

\begin{figure}
    \includegraphics[scale=.63]{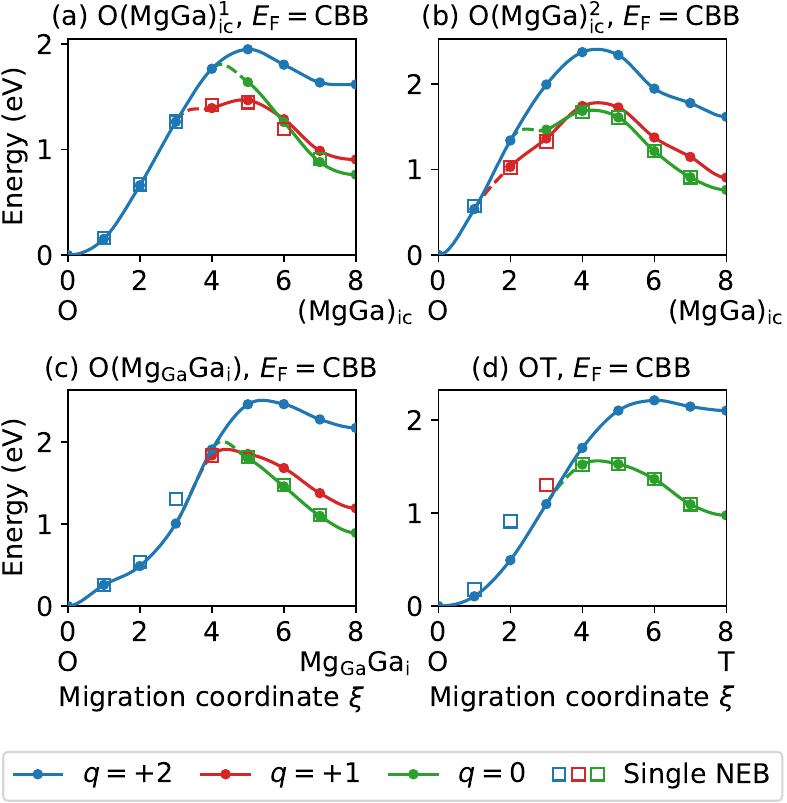}
    \caption{Total energy profiles along the four distinct migration pathways, \textbf{(a)} O\MgGaIC 1, \textbf{(b)} O\MgGaIC 2, \textbf{(c)} O(\MgGaGai), and \textbf{(d)} OT. The energy profile along each pathway is shown for $+2$ (blue), $+1$ (red) and neutral (green) charge states. The abscissa is the migration coordinate $\xi$ representing a particular intermediate structure in multi-dimensional atomic coordinates along the migration path. Relative energy positions among different charge states depend on the Fermi-level ($\EF$) position and in this figure $\EF$ is set to be at the conduction band bottom.
    }\label{fg:rc}
\end{figure}

After identifying the migration pathways described above, we have obtained the energy profiles along the pathways by HSE calculations. At first, we have calculated the energy profiles along
the OO$^z$, OO$^x$ (not shown) and OT [\FIG{fg:rc} (d)] pathways for Mg$^{++}$. The obtained profiles are similar to those reported in the previous work \cite{miceli2017migration}. The calculated migration barriers are 2.02, 1.95 and \SI{2.21}{eV} for the OO$^z$, OO$^x$ and OT, respectively in the present calculations, which are close to the previous values, 2.01, 2.19 and \SI{2.20}{eV}. 
Only for the OO$^x$, the values are different by \SI{0.24}{eV} (10\%). 
We believe that our result is more accurate due to stricter convergence criterion in our calculation which is necessary to determine the transition-state geometry in the plateau-like energy landscape in the case. 

The calculated energy profiles along the pathways from O sites to metastable configurations, \MgGaic, \MgGaGai{} and the T sites, are shown for Mg$^{++}$ in \FIG{fg:rc} as a function of the migration coordinate $\xi$. The coordinate $\xi$ represents a particular intermediate structure in multi-dimensional atomic coordinates along each migration pathway.
Among all pathways for Mg$^{++}$, we have found that the minimum migration energy barrier between O sites is \SI{1.95}{eV} and achieved in the OO$^x$ and O\MgGaIC 1 pathways (\TBL{tb:barrier}). While the obtained minimum value is lower than the previous value by 10\%, the migration barrier is still large compared to that evaluated by the experiments. 

\begin{figure}
    \includegraphics[scale=.63]{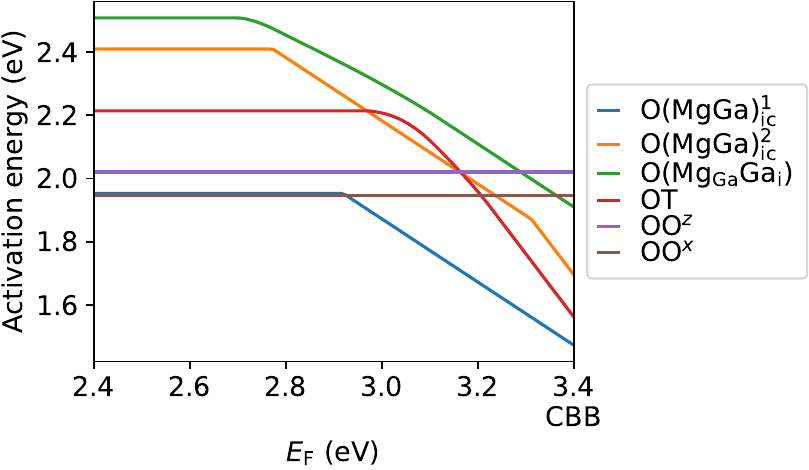}
    \caption{The migration barriers as a function of the Fermi-level position in the gap for distinct migration pathways. All energy barriers are constant below $\text{CBB} - \SI 1{eV}$.}
    \label{fg:ea-ef}
\end{figure}

As stated above, the metastable configurations induce gap states and thus take different charge states, +1 and neutral. We have found that this is also true for the intermediate geometries along the pathways. This finding brings forth a possibility of the migration in which the charge state is varying (recombination enhanced/retarded migration). When $\EF$ is close to CBB, the $+1$ and the neutral metastable states are energetically favorable (\FIG{fg:formation-erg}) and thus the recombination enhanced migration is expected to be important.

To verify this, we have performed a series of NEB calculations with all three charge states. Let us focus on the migration through the O\MgGaIC 1 pathway [\FIG{fg:rc} (a)], in which the migration barrier for Mg$^{++}$ is the lowest, \SI{1.95}{eV}. 
The results from the NEB calculations with $q=+1$ and $0$ are also shown in \FIG{fg:rc} (a). 
When the Mg starts to migrate ($\xi\le 3$), the only possible charge state is $q=+2$ (no gap state). As Mg proceeds ($\xi\ge 4$), the gap state appears and the charge can become $q=+1$ with much reduced formation energy than $q=+2$. 
Near the transition state for the Mg$^{++}$ migration ($\xi \approx 5$), the gap state can accommodate 2 electrons and $q=0$ is also available. 
The relative energies for different charge states depend on the $\EF$ position, as is demonstrated in \FIG{fg:formation-erg}. 
\FIG{fg:rc} shows the results when $\EF$ is at CBB. In this case, the present calculations have unequivocally clarified that Mg starts at the O site as the +2 charge state, successively captures the first and then the second electrons to lower its energy, with the charge state being changed to +1 then neutral, and reaches the \MgGaic{} configuration. The calculated migration barrier is \SI{1.47}{eV} and the electron capture substantially reduces the migration barrier. 

The recombination-enhanced Mg migration clarified above relies on the ansatz that the lowest-energy charge state is reached during the migration. 
This is generally true since the ionic motion is much slower than the electronic transition. 
Yet a particular charge state of defects can be long-lived in semiconductors, e.g., DX center in compound semiconductors \cite{mooney1990deep}.
This is usually associated with structural reconstruction between distinct charge states, where an energy barrier exists. 
In the present calculations, this is already taken into account in the DFT-NEB scheme since there is no abrupt structural changes along the migration pathways.
Another important factor is the peculiar electronic structure of the interstitial Mg. 
It generally induces localized electron states in the band gap near CBB consisting mainly of Ga and N atomic orbitals nearby as shown in \FIG{fg:stable-sites}.
However, at the O site, the interstitial Mg induces no localized gap states but resonant states in the conduction (and also valence) bands. This is presumably due to the high symmetry at the O-site geometry, as in the tetrahedral interstitial site in Si \cite{pantelides1984microscopic,baraff1983theory}.
As a result, at the O site, Mg becomes $q=+2$ with 2 valence electrons transferred to the conduction band. 
Then during the migration towards other metastable geometries with the lower symmetry, we have found that the resonant state shifts downward and emerges in the gap, being ready to capture electrons, especially at the vicinity of separation from the conduction band. No particular energy barriers emerge upon the electron capture, as discussed above.

However, the above NEB calculations with distinct charge states do not include the possible energy barrier resulting from the slightly different local structures of distinct charge states at the same $\xi$.
To clear out this issue, we perform another single NEB calculation connecting the Mg$_{\text O}^{++}$ and \MgGaIC{0} structures by setting $q=+2$, $+1$ and $0$ for images $\xi\le3$, $4\le\xi\le 6$ and $\xi=7$, respectively.
In this way, the structural change along the migration path becomes fully ``continuous''.
The energy profiles obtained by this single NEB is shown by the empty squares in \FIG{fg:rc} (a). It is found that the energy profiles are slightly different among distinct NEB calculations due to local relaxation at different $q$, but the overall energy barriers are essentially identical.

When $\EF$ is lowered by $\Delta\epsilon$ from CBB, the energy profiles for $q=+1$ and $0$ along the migration pathways
are shifted upwards by $\Delta\epsilon$ and $2\Delta\epsilon$, respectively, relative to the energy for $q=+2$, as is derived from \EQ{eq:form-erg}, leading to the increase of the  migration barrier. The migration barrier 
$E_{\text a} (\EF)$ 
at arbitrary $\EF$ position in the gap is calculated from the computed results described above. For each pathway, it is 
\[\textsub Ea (\EF)=\max_{0\le\xi\le8}\left\{\min_{q\in\{+2,+1,0\}} E_q(\xi,\EF)\right\}.\]
Here $E_{q} (\xi, \EF)$ is the formation energy of the intermediate structure $\xi$ with the charge state $q$ at $\EF$. 
The result for the O\MgGaIC 1 pathway is shown by the blue line in \FIG{fg:ea-ef} and we observe a linear decrease of the migration barrier for $\EF$ above $\text{VBT}+\SI{2.9}{eV}$.

We have performed the same calculations for other pathways, the O\MgGaIC 2, O(\MgGaGai), and OT. The results are shown in \FIG{fg:rc} (b)--(d) and \FIG{fg:ea-ef}.
The Mg migration along those pathways is found to exhibit similar behavior as in the O\MgGaIC 1 pathway: i.e. the migration barrier is reduced with increasing $\EF$ due to the recombination enhanced migration.
In the case of OT path, although $q=+1$ does not exist at $T$, we find that the Mg still captures the electrons consecutively similar to other pathways and $q=+1$ is the most stable at intermediate structure $\xi=3$.

For direct migration
through the OO$^z$ and OO$^x$ pathways, no gap state appears during the migration 
and the recombination enhancement of the migration is not observed with the migration barrier being
unaffected by the $\EF$ position (\FIG{fg:ea-ef}).

We summarize
the energy barrier for each 
pathway
at $\EF=\text{VBT}$ and CBB in \TBL{tb:barrier}. At high Fermi level, the lowest energy barrier for migration between O sites is \SI{1.47}{eV}, almost \SI{0.5}{eV} lower than that for low Fermi level. 
There are also multiple pathways with energy barriers well below \SI{2}{eV} and agree with the values obtained from experiments. 
Additionally, it is discovered that when annealing at 
high-pressure N$_2$,
the diffusion of Mg in GaN is largely suppressed \cite{sumida2021effect},
but the reason remains unknown.
Here, we can explain this by considering the intrinsic defects.
The N vacancies in GaN can act as electron donors \cite{lyons2017computationally,miceli2016self}
and raise the Fermi level, assisting the migration of Mg.
Under high N$_2$ ambient pressure, the concentration of N vacancy decreases, and the Mg migration is then suppressed by $\EF$ lowering.

\begin{table}
    \caption{Migration barrier (\si{eV})} of each pathway at $\EF=\text{VBT}$ and CBB\label{tb:barrier}
    \begin{tabular}{cS[table-format=1.2]S[table-format=1.2]}
        \hline\hline
        Pathways & {$E_a(\EF=\text{VBT})$} & {$E_a(\EF=\text{CBB})$} \\
        \hline
        OO$^z$     & 2.02          & 2.02\\
        OO$^x$     & \textbf{1.95} & 1.95\\
        OT         & 2.21          & 1.56\\
        O\MgGaIC 1 & \textbf{1.95} & \textbf{1.47}\\
        O\MgGaIC 2 & 2.41          & 1.70\\
        O(\MgGaGai)  & 2.51          & 1.91\\
        \hline\hline
    \end{tabular}
\end{table}

In conclusion, we have studied the structure of Mg interstitial in GaN and migration pathways and barriers between the (meta)stable configurations from \textit{ab initio} calculations.
In addition to previously reported O and T sites, we have discovered two configurations \MgGaic{} and \MgGaGai{} with formation energy lower than or close to that of T configuration.
Except for Mg at O site which only has $+2$ charge state, all other configurations also admit charge states $+1$ and $0$.
For low Fermi level, the minimum migration energy barrier for Mg$^{++}$ between O sites is \SI{1.95}{eV}.
When Fermi energy is close to CBB, the migration via metastable configurations 
occurs through the
recombination-enhanced mechanism.
The charge state changes from $+2$ at O site to $0$ at metastable sites by capturing two electrons successively during the migration.
This process greatly reduces the migration energy barrier and the minimum energy barrier is reduced to as low as \SI{1.47}{eV} for 
$\EF$ at CBB,
a value consistent with experiments. 

\begin{acknowledgments}
This work is supported by the MEXT programs ``Creation of innovative core technology for power electronics'' (Grant Number JPJ009777), as well as Grant-in-Aid (Kakenhi) for Japan Society for the Promotion of Science (JSPS) Fellows No. 21J20720. 
The computation in this work has been done using the facilities of the Supercomputer Center, the Institute for Solid State Physics, the University of Tokyo.
\end{acknowledgments}

\section*{DATA AVAILABILITY}
The data that support the findings of this study are available from the corresponding author upon reasonable request.

\bibliography{rc.bib}

\end{document}